\documentclass[twocolumn,english,aps,floatfix,superscriptaddress,showpacs,amsfonts,amssymb,superscriptaddress]{revtex4}
\usepackage{color}
\usepackage{color}
\usepackage{graphicx}
\usepackage{amsmath}
\usepackage{latexsym}
\usepackage{epstopdf}
\usepackage{amsmath}
\usepackage{amssymb,amsmath}
\usepackage{multirow}
\usepackage{mathtools}
\newcommand{\beq}{\begin{equation}}
\newcommand{\eeq}{\end{equation}}

\def\ket#1{|#1\rangle}

\newcommand{\eq}{\begin{equation}}
\newcommand{\en}{\end{equation}}
\newcommand{\ear}{\begin{eqnarray}}
\newcommand{\rae}{\end{eqnarray}}

\newcommand{\Tr}{{\rm tr}^{~}}
\newcommand{\rf}[1]{(\ref{#1})}

\def\ket#1{|#1\rangle}

\begin{document}
\title{Generalized mutual informations of quantum critical chains}

\author{F.~C.~Alcaraz}

\affiliation{ Instituto de F\'{\i}sica de S\~{a}o Carlos, Universidade de S\~{a}o Paulo, Caixa Postal 369, 13560-970, S\~{a}o Carlos, SP, Brazil}

\author{M.~A.~Rajabpour}
\affiliation{ Instituto de F\'isica, Universidade Federal Fluminense, Av. Gal. Milton Tavares de Souza s/n, Gragoat\'a, 24210-346, Niter\'oi, RJ, Brazil}

\date{\today{}}

%\date{\today{}}
\begin{abstract}
We study the  generalized mutual information $\tilde{I}_n$ of the ground state of different critical quantum chains. The  generalized mutual information definition 
that we use is based
on the well established concept of the R\'enyi divergence. We calculate this quantity  numerically
for several  distinct quantum   chains having  either discrete 
 $Z(Q)$ symmetries (Q-state Potts model with $Q=2,3,4$ and $Z(Q)$ parafermionic models with $Q=5,6,7,8$ and also
Ashkin-Teller model with different anisotropies) or the 
 $U(1)$  continuous symmetries
(Klein-Gordon field theory, XXZ  and spin-1 Fateev-Zamolodchikov quantum chains with different 
anisotropies). 
For the spin chains these calculations were done by expressing the ground-state wavefunctions in 
two special basis. Our results indicate
 some general behavior for 
  particular ranges of values of the parameter  $n$  that defines 
$\tilde{I}_n$. 
 For a system, with total size $L$ and subsystem sizes $\ell$ and $L-\ell$, 
the   
$\tilde{I}_n$ has a logarithmic leading  behavior   given by $\frac{\tilde{c}_n}{4}\log(\frac{L}{\pi}\sin(\frac{\pi \ell}{L}))$ where 
the coefficient $\tilde{c}_n$  is linearly dependent on the 
central charge $c$ of the underlying conformal field theory (CFT) describing 
the system's  critical properties. 

\end{abstract}
\pacs{11.25.Hf, 03.67.Bg, 89.70.Cf, 75.10.Pq}
\maketitle

\section{Introduction} 

The entanglement entropy, as a tool to detect and classify quantum phase transitions,  has been playing an important role in the last fifteen years (see \cite{EE} and references
therein).
In one dimension, where most of the critical quantum chains are conformal invariant, the entanglement entropy provides a powerful tool to detect, as well to 
calculate, the central charge $c$ of the underlying CFT. For example, for 
 quantum chains, the ground-state entanglement entropy of a subsystem 
formed by contiguous   $\ell$ sites   of an infinite system, 
with respect to the complementary subsystem  has the leading behavior 
% In particular in one dimensional short-range quantum systems the entanglement entropy of the ground state of 
%a subsystem with size $l$
% of an infinite system with respect to the rest is given by 
$S=\frac{c}{3}\ln \ell$ if the system is  critical or $S=\frac{c}{3}\log \xi$, 
 when the system is noncritical 
 with   correlation length $\xi$  \cite{CC2009}. 
%and $c$ is the central charge of the system \cite{CC2009}. i
Although there are 
 plenty of proposals to measure this quantity in the lab \cite{Cardy2011,Demler2012,Zoller2012} the actual experiments were out of reach so far. 
 Strictly speaking
 the central charge of quantum spin chains has never been measured experimentally.  Recently other quantities, that are also  dependent of the central 
charge has been proposed \cite{fidelity,othermeasures}. Among these proposals 
 interesting measures 
 that, from the numerical point of view, are also 
efficient 
 in detecting the phase transitions as well as  the universality 
class of critical behavior, are the Shannon and R\'enyi mutual informations \cite{AR2013,Stephan2013,Stephan2014,AR2014} (see also the related works 
\cite{Wolf2008,Stephan2009,Stephan2010,Oshikawa2010,
 Stephan2011,Um2012,Wai2013}). The R\'enyi mutual information (the exact 
definition will be given in the next section) has a parameter $n$ that recovers 
the Shannon mutual information at the value $n=1$. 
The results derived  in \cite{AR2013,Stephan2013,Stephan2014,AR2014} indicate that the Shannon and R\'enyi mutual informations
 of the ground state of quantum spin chains, 
when expressed in some special local basis, similarly as happens  with 
the Shannon and R\'enyi entanglement entropy,  show a  logarithmic behavior 
with the subsystem's size whose  
 coefficient  depends on the central charge.

 Recently additional  new results  concerning the Shannon and R\'enyi mutual information in quantum systems were obtained,
 see \cite{GR2013,Alet2013,Alet2014a,Alet2014b,Alet2014c}. There are also 
studies of the mutual information in classical two dimensional
 spin systems \cite{Wai2013,wilms,Melko2013,Melko2014,rahmani2013,Rittenberg2014}. It is worth mentioning that
 the Shannon and R\'enyi mutual informations studied in the above papers, as will be defined in the next section, are basis
 dependent quantities. It is  important to  distinguish them from the more known basis independent quantity, namely, the von Neumann mutual information. For recent 
 developments on the calculation of the von Neumann mutual information in thermal equilibrium and non-equilibrium systems see \cite{Eisert2015,Eisler}.

 Most of the results regarding the Shannon and the R\'enyi mutual information, except for the case of harmonic chains, are based on 
 numerical analysis, especially for systems with
 central charge not equal to one. One of the main problems 
in a possible analytical derivation comes from the presence of 
 a  discontinuity at $n=1$ 
 of the R\'enyi mutual information.
This discontinuity prevents the use of the replica trick, which is normally 
a necessary step for the analytical derivation of the Shannon mutual 
information.

In this paper we will
 consider, for many different quantum chains,  another version of the  mutual information, which is also parametrized by a parameter  $n$ that   reduces at $n=1$ to the  Shannon mutual information.
  The motivation for our calculations is two fold. Firstly this definition 
is more appropriate from the point of view of a measure of shared information among parts of a 
system, since it has the expected properties. This will be discussed in the Appendix.  

Secondly, this quantity does not
 show any discontinuity at $n=1$, so it might be a good starting point for 
the analytical calculation of the Shannon mutual information with some sort
 of analytical continuation of the parameter $n$. From now on we will call this new quantity
 generalized mutual information.

Having the above motivations in mind we firstly calculated numerically (using exact diagonalization) the 
generalized mutual information  for several 
critical quantum spin chains. We considered models with $Z(Q)$ symmetries 
like the $Q$-state Potts modes for 
  $Q=2,3$ and $4$, the Z(4) Ashkin-Teller model and the 
 $Z(Q)$ parafermionic models with $Q=5-8$. We then calculated the 
generalized mutual information for quantum critical
 harmonic chains (discrete version of Klein-Gordon field theory) and also 
for quantum spin chains with $U(1)$ symmetry like the XXZ and the spin-1 Fateev-Zamolodchikov quantum chains. 
 
 The structure of the paper is as follows: in the next section we will present the essential definitions of the Shannon and R\'enyi mutual 
 informations as well as generalized mutual information.
  In section three we will present the numerical results
 of the  generalized mutual information for many different critical 
quantum spin chains. Finally in the last section we present our conclusions.

\section{ The generalized mutual informations: definitions} 

Consider the normalized ground state eigenfunction of a quantum spin chain Hamiltonian  $\ket{\psi_G}=\sum_I a_I\ket{I}$, expressed in a particular local basis 
$\ket{I}=\ket{i_1,i_2,\cdots}$, where $i_1,i_2,\cdots$ are the eigenvalues of 
 some local operators  defined on the lattice sites. The R\'enyi entropy is defined as 
\begin{equation} \label{Renyi1}
Sh_n({\cal X}) =\frac{1}{1-n}\ln \sum_I p_I^n,
\end{equation}
where $p_I=|a_I|^2$ is the probability of finding the system in  the 
particular configuration given by $\ket{I}$. The limit $n\to 1$  gives us the Shannon entropy $Sh =-\sum_I p_I \ln p_I$.
Since we are considering  only local basis it is always possible to
decompose the configurations as a combination of the configurations  
 inside and outside of  the subregions as $\ket{I}=\ket{I_AI_{\bar{A}}}$.  One can define the marginal probabilities as
$p_{I_{A}}=\sum_{I_{\bar{A}}} p_{I_{A}I_{\bar{A}}}$ and 
$p_{I_{\bar{A}}}=\sum_{I_A} p_{I_AI_{\bar{A}}}$.

In a previous paper \cite{AR2014} we studied the naive definition of the 
R\'enyi mutual information: 
\begin{equation} \label{Renyi2}
I_n(A,\bar{A})= Sh_n(A)+Sh_n(\bar{A})-Sh_n(A\cup\bar{A}).
\end{equation}
 From now on instead of using $p_{I_AI_{\bar{A}}}$
we will use just $p_I$. 
The known results of the R\'enyi mutual informations of quantum critical 
chains are obtained by using the definition 
\rf{Renyi2}. For special basis, usually the ones where part of the 
Hamiltonian is diagonal (see \cite{AR2014}), the definition \rf{Renyi2} 
for the R\'enyi mutual information gives us a logarithmic behavior with 
the subsystem size, for arbitrary values of $n$. However, as observed 
numerically for several quantum chains 
(see \cite{Stephan2014,AR2014,Stephan2009}), it shows a discontinuity 
at $n=1$, that forbids  the use of large-$n$ analysis to obtain the most 
interesting case where $n=1$, namely   the standard Shannon mutual 
information. 
Although the  definition \rf{Renyi2} has its own uses it is not the one which 
normally has been considered in information sciences. For example $I_n$ for $n\neq 1$ is not necessarily a positive function, a property
that we naturally expect to be hold for the mutual informations.  In 
this paper  we consider  a definition that  is common in 
information sciences \cite{Principe}. 
The generalized mutual information with the desired properties, 
as a measure of shared information  (see Appendix),  is defined as
\cite{Principe}:
\begin{eqnarray}\label{Renyi new}
\tilde{I}_n(A,\bar{A})=\frac{1}{n-1}\ln\sum_I\frac{p_I^n}{p_{I_{\bar{A}}}^{n-1}p_{I_{A}}^{n-1}},
\end{eqnarray}
where $p_{I_{A}}$ and $p_{I_{\bar{A}}}$, as before,  are the probabilities 
that the subsystems are independently in the configurations 
$\ket{I_A}$ and $\ket{I_{\bar{A}}}$ that forms the configuration $\ket{I}$ 
that occurs with probability $p_I$.

Hereafter  $L$ will represent the size of the whole system and $\ell$ and 
$L-\ell$   the sizes of the subsystems. With this  new notation
one can write $\tilde{I}_n(A,\bar{A})$ as $\tilde{I}_n(\ell,L-\ell)$. This definition of the generalized mutual information comes from the natural 
extension of the relative entropy to the R\'enyi case and  measures the distance of the full distribution from the product
of two  independent distributions.
In the limit
$n\to1$ one  easily recovers the Shannon mutual information $\tilde{I}_1(l,L-\ell)=Sh(\ell)+Sh(L-\ell)-Sh(L)$, where 
$Sh =-\sum_I p_I \ln p_I$ is the standard Shannon entropy. One of the important properties of  $\tilde{I}_n$, that is not shared by $I_n$, is  its  nondecreasing behavior as a function  of $n$ 
 (see Appendix). 
 Our 
calculations 
 for  a set of distinct quantum spin chains will be done 
numerically, since up to our knowledge an analytical method to consider 
these quantum chains is still  missing.

\section{The generalized mutual information in quantum  chains}

In this section we will numerically calculate the ground-state generalized mutual information   of  two series of critical
quantum spin chains with slightly different structure. In the first part we will calculate the generalized mutual information for systems with
discrete symmetries such as
the $Q$-state Potts models 
 with $Q=2,3$ and $4$, the Ashkin-Teller model and the 
parafermionic $Z(Q)$-quantum spin chain \cite{f-z} 
for the values of $Q=5,6,7$ and $8$. In the second part we will calculate the generalized mutual information for  systems with
$U(1)$ symmetry such as the Klein-Gordon field theory, the XXZ model and the Fateev-Zamolodchikov model with 
different
values of their anisotropy parameters.
%parei aqui

\subsection{The generalized mutual information in quantum chains with discrete symmetries}

In this subsection we will study the  generalized mutual information of the ground state of different critical
spin chains with $Z(Q)$ discrete symmetries. 
The results we present were obtained by expressing the ground-state 
wavefunction  in two specific basis where  the systems show some universal properties.
\begin{figure} [htb] \label{fig1}
\centering
\includegraphics[width=0.35\textwidth]{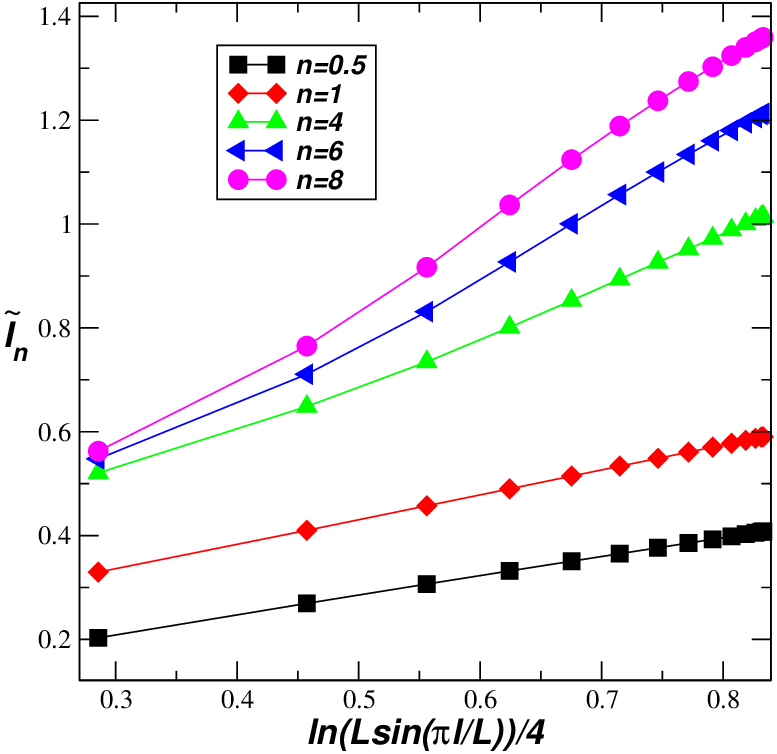}
\caption{  The generalized mutual information $\tilde{I}_n(\ell, L-\ell)$ of the $L=28$ sites periodic Ising 
quantum chain, as a function of 
 $\ln(L\sin(\frac{\pi \ell}{L}))/4$. The ground-state wavefunction is in
 the  basis where the  matrices $S_i$ are  diagonal ($S$ basis).}
\end{figure}
\newline

\subsubsection{The generalized mutual information of the quantum $Q$-state Potts model and  the quantum Ashkin-Teller model}

Our results show that  the Q-state Potts model and the Ashkin-Teller model 
share a  similar behavior. For this reason  we  discuss them together.
The critical $Q$-state Potts model in a periodic lattice is defined by the Hamiltonian \cite{Wu1982}

\begin{eqnarray}\label{Potts Hamiltonian}
H_Q=-\sum_{i=1}^L\sum_{k=1}^{Q-1}(S_i^kS_{i+1}^{Q-k}+ R_i^k),
\end{eqnarray}
where $S_i$ and $R_i$ are $Q\times Q$ matrices satisfying the following $Z(Q)$ algebra: 
$[R_i,R_j]=[S_i,S_j]=[S_i,R_j]=0$ for $i\neq j$ and
$S_jR_j=e^{i\frac{2\pi}{Q}}R_jS_j$ and $R_i^Q=S_i^Q=1$.
The model has its  critical behavior  governed by a CFT with
 central charge $c=1-\frac{6}{m(m+1)}$ where
$\sqrt{Q}=2\cos(\frac{\pi}{m+1})$. The $Q=2$ Potts chain is just the 
standard Ising quantum chain. 
\begin{figure} [htb] \label{fig2}
\centering
\includegraphics[width=0.35\textwidth]{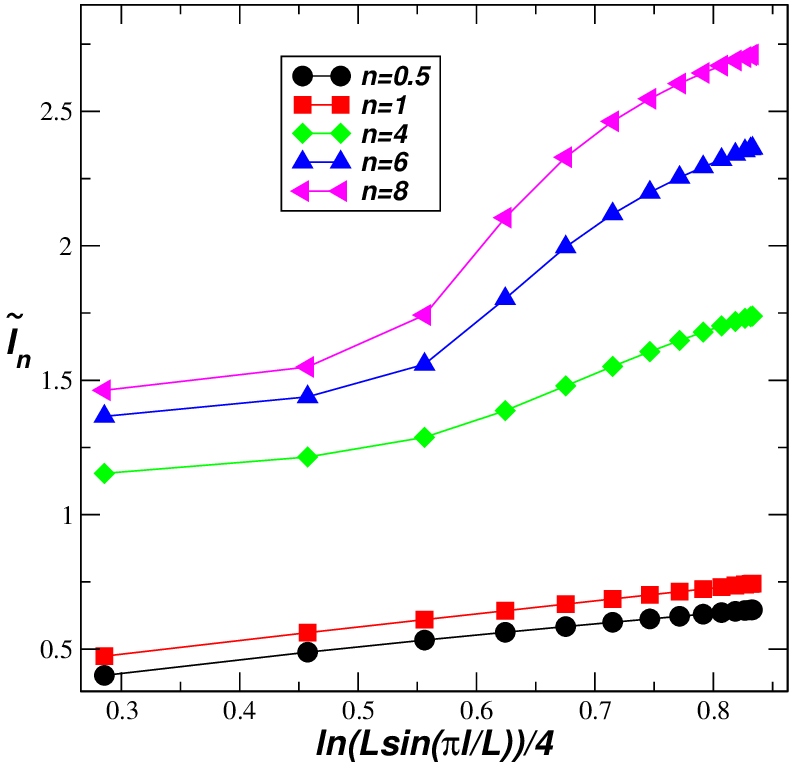}
\caption{ The generalized mutual information $\tilde{I}_n(\ell, L-\ell)$ of the $L=28$ sites periodic Ising 
quantum chain, as a function of 
 $\ln(L\sin(\frac{\pi \ell}{L}))/4$. 
The ground-state wavefunction is in   the  basis where the matrices $R_i$ are diagonal ($R$ basis).}
\end{figure}
The Ashkin-Teller model has a $Z(2)\otimes Z(2)$ symmetry  and  a Hamiltonian  given by:
\begin{equation}
H = -\sum_{i=1}^L\Big{(} S_i S_{i+1}^3 +S_i^3 S_{i+1} +\Delta S_i^2S_{i+1}^2  +R_i +R_i^3 
+\Delta R_i^2\Big{)},
\end{equation}
where $S_i$ and $R_i$  are the same  matrices introduced  in the  
 $Q=4$  Potts model. 
The model is critical and conformal invariant for $-1<\Delta\leq 1$ with the central charge $c=1$. It is worth mentioning
that at $\Delta = 1$ we recover the  $Q=4$ Potts model and at $\Delta =0$ the 
 model  is equivalent to two decoupled Ising models.
 
In the paper  \cite{AR2014} we already showed that the  
 Shannon and R\'enyi mutual informations, as defined in \rf{Renyi2}, 
  are basis dependent. In other words one can get quite distinct different finite-size scaling behaviors by considering 
different basis. Surprisingly in some particular basis, that we called  conformal basis, the results shows some universality. 
For example, 
the results for the $Q$-state Potts model and for the Ashkin-Teller model in the basis where 
the matrices $R_i$ or the matrices $S_i$  are diagonal  are 
the same, and follow the asymptotic behavior
\begin{eqnarray}\label{Shannon MI} 
I_n(\ell,L-\ell)=\frac{c_n}{4}\ln(\frac{L}{\pi}\sin(\frac{\pi \ell}{L}))+...,
\end{eqnarray}
with
\begin{equation}\label{c-n conformal Potts}
c_n=c\left\{
\begin{array}{c l}      
    1, & n=1\\
\frac{n}{n-1}, & n>1.5 
\end{array}\right. .
\end{equation}
We should mention  that in \cite{Stephan2014}, based on numerical results, 
it was claimed that for $n=1$ the coefficient $c_1$ might not be exactly equal to the central charge. 
As it was discussed  in \cite{Stephan2014,AR2014} it is quite likely 
that $I_n$ is not a continuous function around $n=1$ and so any attempt 
to do  
the replica trick using this definition of
R\'enyi mutual information will be useless. 
This makes the analytical calculation a challenge. This is an additional 
reason to examine  the behavior of $\tilde{I}_n$, besides being the 
correct extension, from the point of view of a measure of 
shared information. Having this in mind we calculated the $\tilde{I}_n$ for 
$Q=2,3$ and $Q=4$ Potts chains and for the Ashkin-Teller model in the $R$ and the $S$ basis. We 
 found that in some regimes of variation of the parameter $n$  one can fit the data nicely to 
\begin{eqnarray}\label{Renyi MI Potts} 
\tilde{I}_n(\ell,L-\ell)=\frac{\tilde{c}_n}{4}\log(\frac{L}{\pi}\sin(\frac{\pi \ell}{L}))+...,
\end{eqnarray}
being $\tilde{c}_n$  a monotonically nondecreasing function of $n$, consistent with what we expect for the  mutual information, since it is a good measure of shared information  (see the Appendix).

\begin{figure} [htb] \label{fig3}
\centering
\includegraphics[width=0.35\textwidth]{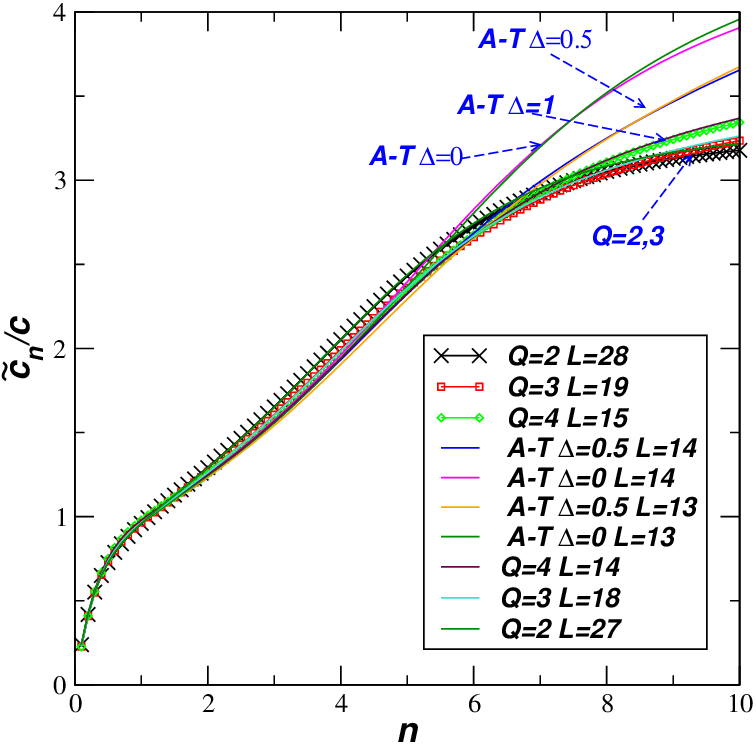}
\caption{The ratio $\tilde{c}_n/c$ of the coefficient of the logarithm in  the equation (\ref{Renyi MI Potts}) and the central charge $c$ for the $Q$-state Potts model 
with $Q=2,3$ and 4,  and for the Ashkin-Teller model (A-T) with different anisotropies $\Delta$. The Ashkin-Teller model at the isotropic point ($\Delta=1$) is equivalent to the 
$4$-state Potts model. The ground-state wavefunctions are in the basis where the $S_i$ matrices are 
diagonal. The lattice sizes of the models are shown and the coefficients $\tilde{c}_n$ were estimated  by using the subsystem 
sizes $\ell=3,5,...,{\mbox{Int}}[L/2]$.} 
\end{figure}

Here we summarize the results for the $Q$-state Potts  and Ashkin-Teller quantum chains:
\begin{enumerate}
 \item The results in general depend on the basis we choose to express the 
ground-state wavefunction.
 
\item The  generalized mutual  information follows  (\ref{Renyi MI Potts})  in the $S$ and $R$ basis but with different coefficients 
for different basis. 
To illustrate the logarithmic behavior we show in Fig.~1 and Fig.~2 
the mutual 
information $\tilde{I}_n$ for the Ising model ($Q=2$) with $L=28$ sites and ground-state 
eigenfunctions  in the $S$ and $R$ basis, respectively.
 We  see, from these figures,  that for subsystem 
sizes $\ell \geq 3$ we have the logarithmic behavior 
given by \rf{Renyi MI Potts} 
up to $n\approx 8$ in 
the $S$-basis and $n\approx 4$ in the $R$-basis. As we can see our results 
does not exclude the existence of some relevant $\ell$-dependent terms in 
\rf{Renyi MI Potts} for  large values of $n$.
\begin{figure} [htb] \label{fig4}
\centering
\includegraphics[width=0.35\textwidth]{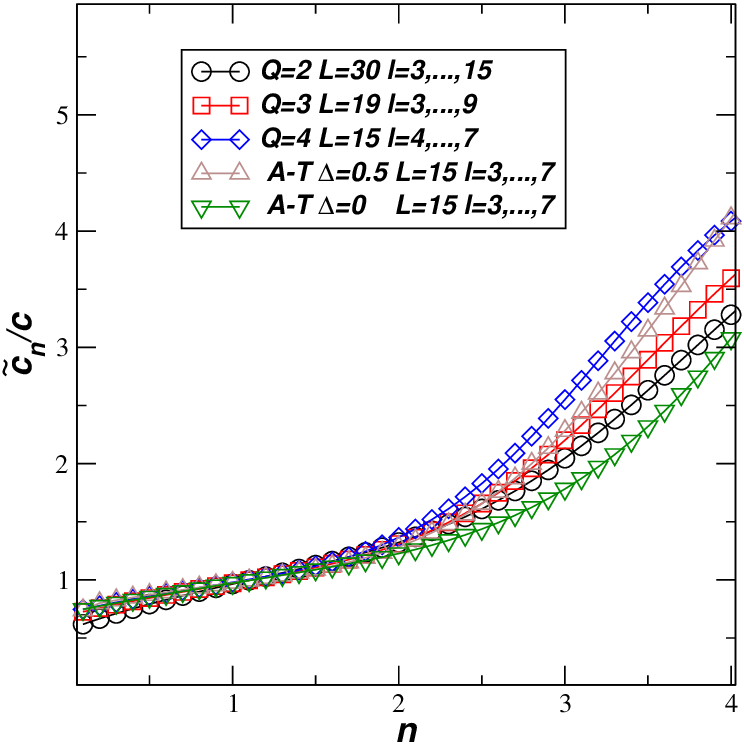}
\caption{Same as Fig.~3, but with the ground-state wavefunctions of the quantum spin Hamiltonians expressed in the basis where the matrices $R_i$ are diagonal.
The lattice sizes of the  models are shown in the figure, as well as the subsystems  sizes $\ell$ 
used to estimate  $\tilde{c}_n$.}
\end{figure}

\item The coefficient of the logarithm $\tilde{c}_n$ in \rf{Renyi MI Potts}  
is a continuous monotonically non-decreasing function of $n$ and 
it follows the following formula in the $S$ basis:
\begin{eqnarray}\label{coefficient} 
\tilde{c}_n=c f(n),\quad \mbox{with} \quad f(1) =1,
\end{eqnarray}
where $c$ is the central charge and $f(n)$ seems to be  a continuous universal function independent of the model, as we can  see in Fig.~3. In the case of
the Ashkin-Teller model the results start to deviate around $n=6$ from the ones
obtained for  the Potts models. As we can see in Fig.~3, the deviation point is   dependent on 
the anisotropy parameter $\Delta$ of the model.
\item In the case of the $R$ basis , as one can see in Fig.~4, equation  (\ref{coefficient}) is still valid  for values of $n$ up to $\sim $4. However  the function $f(n)$ is distinct from the one obtained in the $S$ basis.
 As shown in Fig.~ 4, up to $n=2$ the form of the function $f(n)$ seems 
to be also independent 
of the model. This figure also shows that the Ashkin-Teller model has
stronger deviations in this basis, as compared with the results obtained in the 
 $S$ basis. 
%insertion for the point 4 of referee I
In order to better see  the difference  of  the coefficients $\tilde{c}_n$ in the
$S$ and $R$ basis, we present in  Fig.~5 the data of Figs.~3 and ~4 for the $Q=2,3$ and 4 state Potts models.
\begin{figure} [htb] \label{fig5}
\centering
\includegraphics[width=0.35\textwidth]{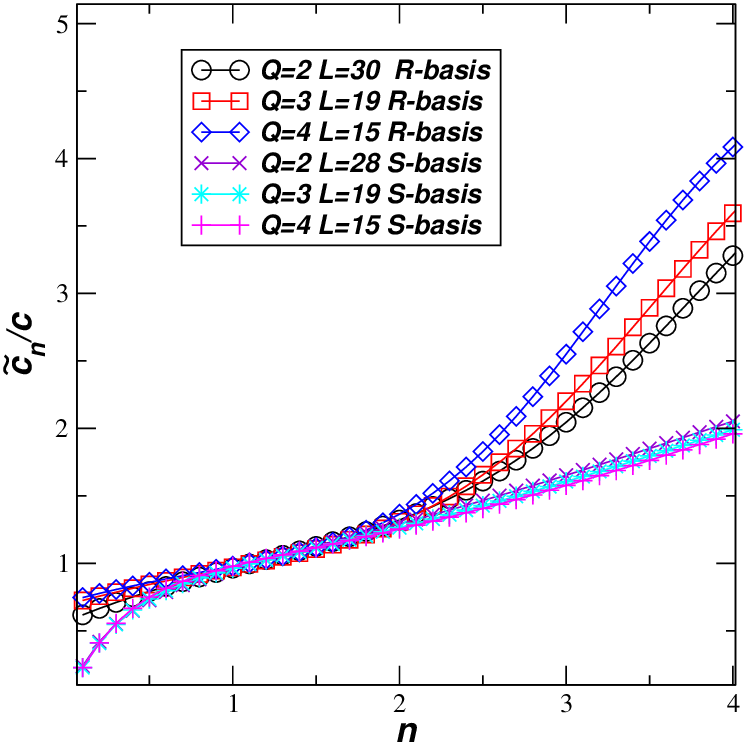}
\caption{The values of the ratios $\tilde{c}_n/c$ of Figs.~3 and ~4 for the
$Q=2,3$ and 4-state Potts model are shown in the same figure, for comparison.}
\end{figure}
%end of insertion of point 4 of referee I
\item The coefficient of the logarithm in the $S$ basis always goes to zero 
as $n\to 0$, differently from the $R$ basis where it approaches to a non-trivial number. This simply means that probably in the continuum limit
all the probabilities in the $S$ basis are positive but in the $R$ basis some of them are zero. For the definition of the $n=0$ case see the Appendix.
\end{enumerate}
Our numerical results indicate  that  $\tilde{c}_n$ is a continuous function of $n$ around $n=1$. This means that $\tilde{I}_n$ should be a   
 continuous function with respect to $n$ and so it is a better candidate 
to be used in 
techniques exploring the analytical continuation of the value $n$, as happens 
for example  
in 
the replica trick.
 However, the appropriate 
 technique that may be used is still unclear to us.

%start the insertion of point 3 of referee I
It is important to mention that the results obtained for the ratio $\tilde{c}_n/c$ in this section 
(Fig.~3) and in the subsequent ones (Figs.~8,~9,~11 and ~13) are based on the linear fit with the $\ln[L\sin(\ell\pi/L)]$ dependence. These
fittings were done by choosing a set of subsystems sizes. In all the presented
figures we only depict  results where a small variation of the number of subsystem sizes gives us estimated 
values of $\tilde{c}_n$ that differs a few percent. As an example we consider the fittings obtained from the data of
Figs.~1 and ~2
for the Ising model with $L=28$ sites and ground-state eigenfunction in the
$S$ and $R$ basis, respectively. This is shown in Fig.~6. As we can see, while for the $S$ basis the fitting is reasonable up to $n=8$ in the $R$ basis we do
not have reliable results for $n>4$.
\begin{figure} [htb] \label{fig6}
\centering
\includegraphics[width=0.35\textwidth]{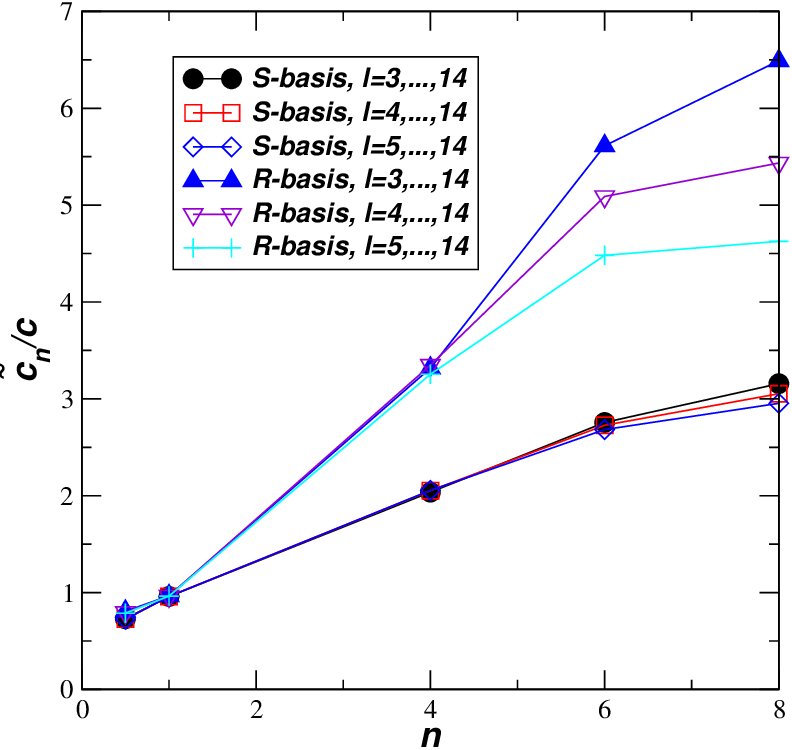}
\caption{The values of $\tilde{c}_n/c$ obtained from the data of Figs.~1 and
~2 for the Ising quantum chain with $L=28$ sites and eigenfunction expressed
in $S$ and $R$ basis.}
\end{figure}
%end of insertion of point 3 of referee I

\subsubsection{The generalized mutual information in the parafermionic 
\\$Z(Q)$-quantum spin chains}
\begin{figure} [htb] \label{fig7}
\centering
\includegraphics[width=0.35\textwidth]{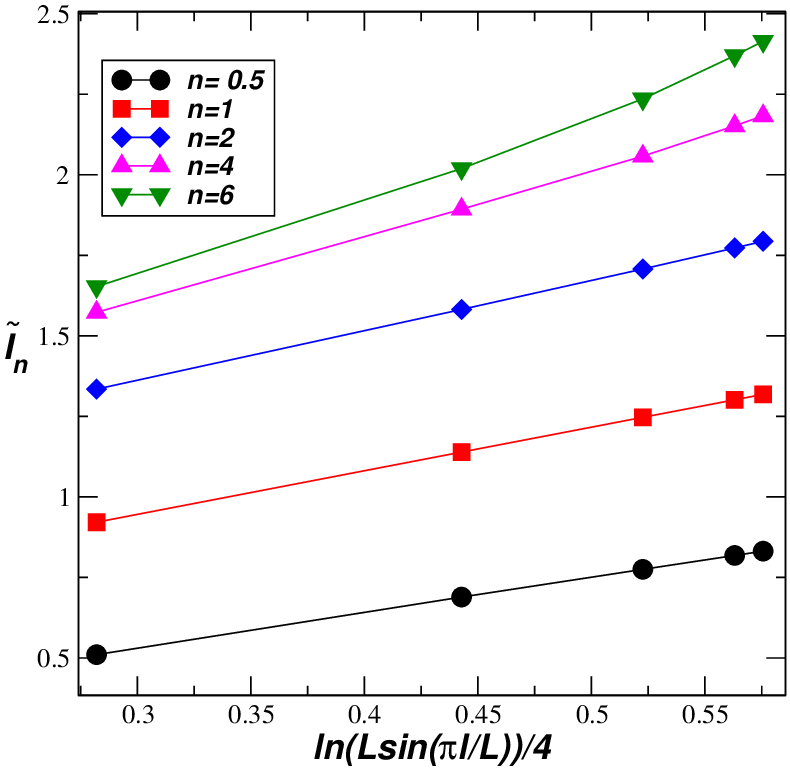}
\caption{ The generalized mutual information $\tilde{I}_n(\ell, L-\ell)$ of the $L=10$ sites periodic $Z(7)$-parafermionic  
quantum chain, as a function of 
 $\ln(L\sin(\frac{\pi \ell}{L}))/4$. The ground-state wavefunction is in   the  basis where the $S_i$ matrices are  diagonal ($S$ basis).}
\end{figure}
In this subsection we consider the generalized mutual information for 
some critical spin chains with discrete $Z(Q)$ symmetry and central charge  bigger than one. 
The quantum chains  we consider are  the parafermionic $Z(Q)$-quantum spin 
chains \cite{f-z} with Hamiltonian  given by \cite{lima,alca}
\begin{equation} \label{zn}
H=-\sum_{i=1}^L\sum_{k=1}^{Q-1} (S_i^kS_{i+1}^{Q-k} + R_i^k)/\sin(\pi k/Q),
\end{equation}
where again $S_i$ and $R_i$ are the $Q\times Q$ matrices that appeared in (\ref{Potts Hamiltonian}). This model is critical and conformal invariant with a central charge 
$c=2(Q-1)/(Q+2)$. For the case where $Q=2$ and $Q=3$ we recover the Ising 
and 3-state Potts model, and for the case where $Q=4$ we obtain 
the Ashkin-Teller model with the anisotropy value $\Delta= \frac{\sqrt{2}}{2}$.
We have considered the models 
with  $Q=5,6,7$ and $8$ and the ground-state wavefunctions  expressed in the 
$S$ or $R$ basis. The results for the several values of $Q$ are shown in 
Figs.~7,~8 and ~9.
To illustrate the logarithmic dependence with the 
subsystem size $\ell$ we show in 
Fig.~7 $\tilde{I}_n(\ell,L-\ell)$, as a function of $\ln(L\sin(\pi\ell/L))/4$ for the $Z(7)$ parafermionic 
quantum chain with $L=10$ sites, with the ground-state wavefunction expressed  in the $S$ basis. In Figs.~8 and ~9 we show the ratio  $\tilde{c}_n/c$ of the logarithmic coefficient of 
\rf{Renyi MI Potts} with the central charge $c$ for the $Z(Q)$-parafermionic 
models with ground-state wavefunction in the $S$ and $R$ basis, 
respectively.
 The maximum lattice sizes we used for the $Z(Q)$-parafermionic models are $L=12,11,10$ and $9$ for $Q=5,6,7$ and 8, respectively.
The results we obtained are very similar to the ones  we already discussed 
in the previous case of the $Q$-state Potts models. All the five properties 
that we discussed in that subsection are equally valid also for the 
$Z(Q)$-parafermionic models.
By comparing the results of Figs.~8 and ~9 with Figs.~3 and ~4  we observe 
 that the function 
$f(n)$ in \rf{coefficient} are quite similar for the two set of models, 
at least for values of $n$ up to $\sim 6$.  Probably the 
matching of these curves is not perfect due to the small system sizes we 
consider, specially for $Q>4$.

\begin{figure} [htb] \label{fig8}
\centering
\includegraphics[width=0.35\textwidth]{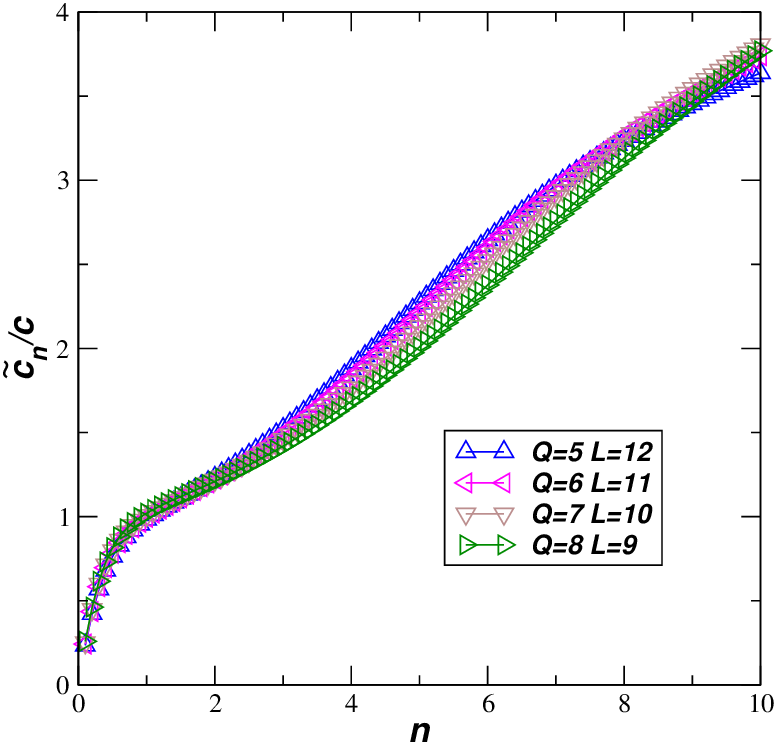}
\caption{The ratio $\tilde{c}_n/c$ of the coefficient of the logarithm in  equation (\ref{Renyi MI Potts}) and the central charge $c$ for the $Z(Q)$-parafermionic  models 
with $Q=5,6,7$ and 8.
 The ground-states are in the basis where the $S_i$ matrices are 
diagonal. The lattice sizes of the models are shown in the figure and the coefficients $\tilde{c}_n$ were estimated  by using the subsystem 
sizes $\ell=3,5,...,{\mbox{Int}}[L/2]$.} 
\end{figure}

\subsection{The generalized mutual information of quantum chains with continuous symmetries}

In this section we consider the  generalized mutual information of  
critical  chains having a  continuous $U(1)$ symmetry.
 We studied a set of  coupled harmonic oscillators which gives a  discrete version of Klein-Gordon field theory as well as the  spin-1/2  XXZ 
and the spin-1 Fateev-Zamolodchikov quantum chains.  
The last two models are interesting since, like the Ashkin-Teller model, they have an anisotropy that gives us  a critical line of continuously varying critical 
exponents but with a fixed central charge.
\begin{figure} [htb] \label{fig9}
\centering
\includegraphics[width=0.35\textwidth]{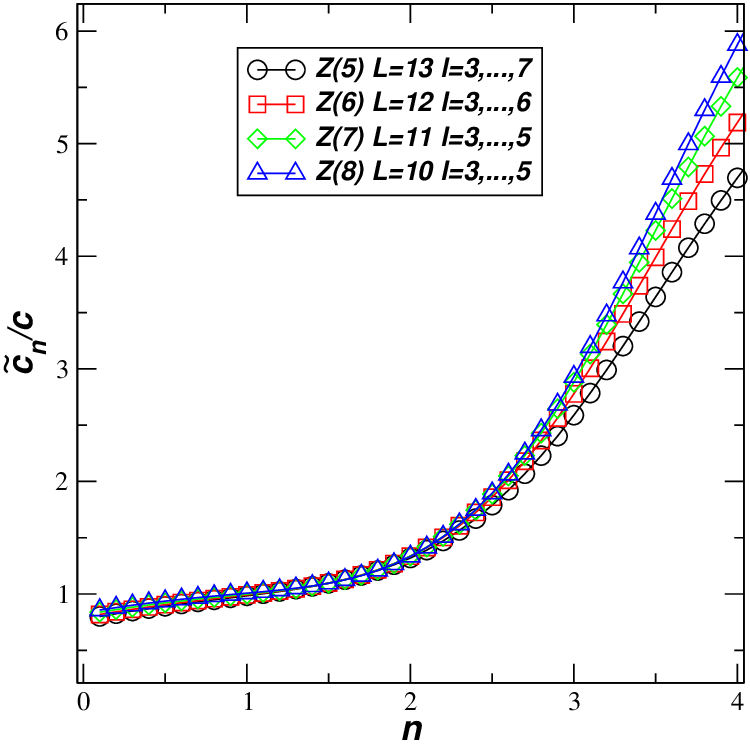}
\caption{Same as Fig.~8, but with the ground-state wavefunction of the quantum spin Hamiltonians expressed in the basis where the matrices $R_i$ are diagonal.
The lattice sizes of the  models, as well as the subsystems  sizes $\ell$ used  
to estimate  $\tilde{c}_n$ are shown.}
\end{figure}
\subsubsection{ The generalized mutual information in quantum harmonic chains} 

In this subsection we will first consider the  generalized mutual information of the ground state of a system of generic coupled harmonic oscillators. 
Then at the very end we will confine ourselves
to the simple case where we have only the  nonzero couplings at the next-nearest sites, that in the continuum limit 
 gives us the  Klein-Gordon field theory. 

Consider the Hamiltonian of $L$-coupled harmonic 
oscillators, with coordinates $\phi_1,\ldots,\phi_L$ and conjugated momenta 
$\pi_1,\ldots,\pi_L$:

\begin{equation}\label{harmonicOsc}
\mathcal{H}=\frac{1}{2}\sum_{n=1}^{L}\pi_n^2+\frac{1}{2} \sum_{n,n^\prime=1}^{L}\phi_{n} K_{nn^\prime}\phi_{n^\prime}~.
\end{equation}
The ground state of the above Hamiltonian has the following form
\begin{equation}\label{GroundSwave}
\Psi_0=(\frac{\det K^{1/2}}{\pi^L})^{\frac{1}{4}} e^{-\frac{1}{2}<\phi|K^{1/2}|\phi>}.
\end{equation}
For the general Hamiltonian  \rf{harmonicOsc}, 
one can calculate the two point correlators $X_A=\Tr(\rho_A \phi_i \phi_j)$ and 
$P_A=\Tr(\rho_A \pi_i \pi_j)$ using the $K$ matrix defined in \rf{harmonicOsc}.
 The squared root of this matrix, as well as its inverse, can be split up up  into 
coordinates of the subsystems $A$ (size $\ell$) and $\bar{A}$ (size $L-\ell$), 
i. e., 
\begin{eqnarray}\label{X_A P_A}
 K^{-1/2}=
\begin{pmatrix}
  X_{A} & X_{A\bar{A}} \\
  X^{ T}_{A\bar{A}} & X_{\bar{A}} 
 \end{pmatrix}, \hspace{1cm} K^{1/2}=
\begin{pmatrix}
  P_{A} & P_{A\bar{A}} \\
  P^{ T}_{A\bar{A}} & P_{\bar{A}} 
 \end{pmatrix}.\nonumber
\end{eqnarray}
Here we chose the couplings so that we always keep the equalities 
 $P^{ T}_{A\bar{A}}=P_{A\bar{A}}$ and
$X^{ T}_{A\bar{A}}=X_{A\bar{A}}$. The spectra of the  matrix $2C = \sqrt{X_AP_A}$, can be used to calculate              
 the R\'enyi entanglement entropy (see \cite{Casini2009} and references therein)  as

\begin{eqnarray} \label{Renyi from corr} 
 S_n(\ell,L-\ell) &=& \frac{1}{n-1}\Tr \left[ \ln\left((C+\frac{1}{2})^n-(C-\frac{1}{2})^n\right) \right]. \nonumber
\end{eqnarray}
In this formulation  we only need the correlators inside the region $A$. Note 
that the above quantity is   basis independent and   is considered  as an   
usual measure of the quantum entanglement. 
Here we need to introduce this quantity just for later use. To calculate
the generalized mutual information for a system of coupled harmonic oscillators one first needs to fix the basis. Here we
work in the position coordinate basis, however all the results are valid also in the momentum basis. One should notice that the same is not true
if one works in a generic basis obtained through canonical transformations from the position or  momentum basis.
In order to calculate $\tilde{I}_n$
first we find $p(\Phi_{A})$ and $p(\Phi_{\bar{A}})$ as

\begin{eqnarray}\label{subsystem probabilities1} 
p(\Phi_{A})=\sqrt{\frac{\det \tilde{P}_A}{\pi^\ell}}e^{-\Phi_A \tilde{P}_A \Phi_A},\\
\label{subsystem probabilities2}
p(\Phi_{\bar{A}})=\sqrt{\frac{\det \tilde{P}_{\bar{A}}}{\pi^{L-\ell}}}e^{-\Phi_{\bar{A}} \tilde{P}_{\bar{A}} \Phi_{\bar{A}}}, 
\end{eqnarray}
where $\tilde{P}_A=P_A-P_{A\bar{A}}(P_{\bar{A}})^{-1}P^T_{A\bar{A}}$ and $\tilde{P}_{\bar{A}}=P_{\bar{A}}-P^T_{A\bar{A}}(P_A)^{-1}P_{\bar{A}A}$. 
Since $\phi$ takes  continuum values one needs to consider the 
integral version of the equation \rf{Renyi new} as follows:
\begin{eqnarray}\label{New renyi continuum} 
 \tilde{I}_n=\frac{1}{n-1}\ln\int\mathcal{D}\Phi \frac{p^n(\Phi)}{p^{n-1}(\Phi_A)p^{n-1}(\Phi_{\bar{A}})},
\end{eqnarray}
where $p(\Phi)=|\Psi_0|^2$. Plugging Eqs. (\ref{GroundSwave}), (\ref{subsystem probabilities1})
 (\ref{subsystem probabilities2}) in the equation (\ref{New renyi continuum}) and performing the Gaussian integral one
 can derive 
 the generalized mutual information 
\begin{eqnarray}\label{Renyi MI HO1} 
&&\tilde{I}_n=\frac{1}{2}\ln \left(\frac{\det K^{\frac{1}{2}}}{\det \tilde{P}_A\det\tilde{P}_{\bar{A}}}\right)\hspace{3.5cm}\nonumber\\&&-
\frac{1}{2(n-1)}\ln\left(\frac{\det\left(nK^{1/2}-(n-1)\begin{pmatrix}
  \tilde{P}_A & 0 \\
  0 & \tilde{P}_{\bar{A}} 
 \end{pmatrix}\right)}{\det K^{1/2}}\right).\nonumber
\end{eqnarray}
The following determinant formulas
\begin{eqnarray}
\det(\tilde{P}_A) \det P_{\bar{A}} =\det K^{1/2},\\
\det(\tilde{P}_{\bar{A}}) \det P_A =\det K^{1/2},\\
\det P_{\bar{A}} \det K^{-1/2}= \det X_A,\\
\det P_A \det K^{-1/2}= \det X_{\bar{A}},
\end{eqnarray}
allow us to   write
\begin{eqnarray}\label{Renyi MI HOC} 
\tilde{I}_n(\ell,L-\ell)&=&S_2(\ell,L-\ell)\nonumber \\ &-&\frac{1}{2(n-1)}\ln\det(n+(1-n)T),
\end{eqnarray}
where 
\begin{eqnarray}\label{Renyi MI HO2} 
T=\begin{pmatrix}
  X_A\tilde{P}_A & X_{A\bar{A}}\tilde{P}_{\bar{A}} \\
  X^T_{A\bar{A}}\tilde{P}_A & X_{\bar{A}}\tilde{P}_{\bar{A}}  
 \end{pmatrix}=
 \begin{pmatrix}
  1 & X_{A\bar{A}}\tilde{P}_{\bar{A}} \\
  X^T_{A\bar{A}}\tilde{P}_A & 1  
 \end{pmatrix}. 
 \end{eqnarray}
 There is an important remark that we should mention: in principle Eq.~(\ref{Renyi MI HOC}) makes sense only if   $n+(1-n)T$ is
 a symmetric positive definite matrix. If we start with a 
symmetric positive definite matrix $K^{1/2}$ this is already warrantied 
 for $0<n<1$ but for 
 $n>1$ one needs to check its validity. This will be an important point when we study the short-range coupled harmonic oscillators.
  Finally one can write
 \begin{eqnarray}\label{Renyi MI HO main} 
\tilde{I}_n(\ell,L-\ell)=S_2(\ell,L-\ell)+\tilde{M}_n(\ell,L-\ell)=S_2(\ell,L-\ell)\nonumber \\
-\frac{1}{2(n-1)}\ln\det(1-(1-n)^2X^T_{A\bar{A}}\tilde{P}_AX_{A\bar{A}}\tilde{P}_{\bar{A}} )\nonumber
\end{eqnarray}
 where $\tilde{M}_n(\ell,L-\ell)$ is the only $n$ dependent part. We notice here that by changing $n$ to $2-n$ we just change the sign of the second term, i. e.,
  $\tilde{M}_{2-n}(\ell,L-\ell)=-\tilde{M}_n(\ell,L-\ell)$.
 
 \begin{figure} [htb] \label{fig10}
\centering
\includegraphics[width=0.35\textwidth]{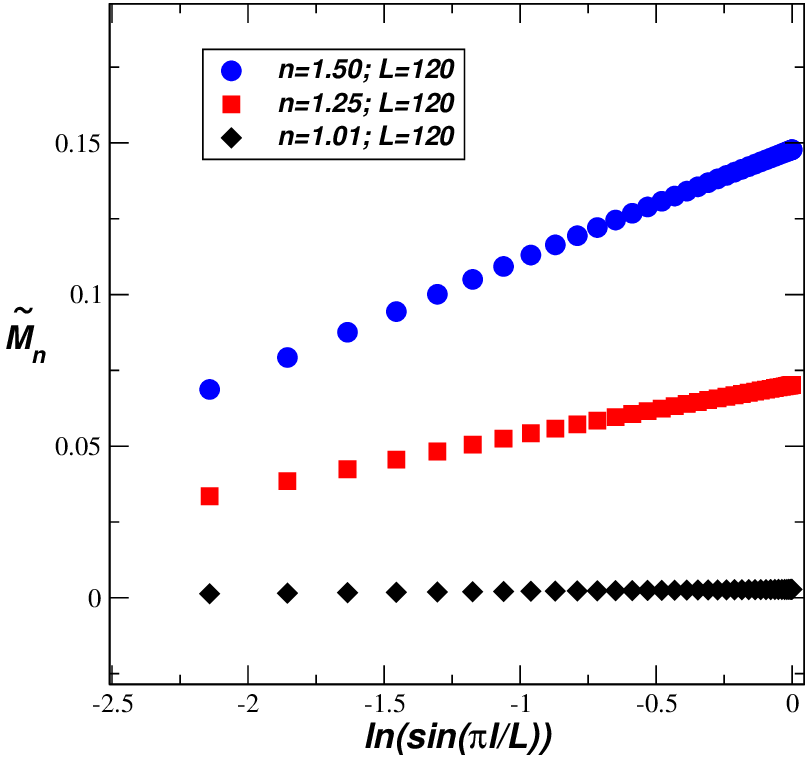}
\caption{ The second term  in the equation (\ref{Renyi MI HO main}), $\tilde{M}_n(\ell,L-\ell)$,
 as a function of 
 $\ln(L\sin(\frac{\pi \ell}{L}))$  for periodic quantum harmonic chain with $L=120$ sites.}
\end{figure}
 
 When $n\to 1$ the second term vanishes and we recover the result of \cite{AR2013} 
\begin{eqnarray}\label{Shannon MI HO1} 
\tilde{I}_1(\ell,L-\ell)=S_2(\ell,L-\ell).
\end{eqnarray}
For massless Klein-Gordon theory the above result in one dimension gives, 
as a consequence  the well known result for the R\'enyi entanglement 
entropy \cite{Casini2009,Cramers2006}, 
\begin{eqnarray}\label{Shannon MI HO} 
\tilde{I}_1(\ell,L-\ell)=\frac{1}{4}\ln(\frac{L}{\pi}\sin(\frac{\pi \ell}{L}))+...,
\end{eqnarray}
where the dots are the subleading terms. Our numerical analyses indicate that for short-range quantum harmonic oscillators the matrix $n+(1-n)T$ is 
symmetric positive definite  up to just $n=n_c=2$\footnote{For finite size system $n_c$ is not exactly equal to $2$, however, by increasing the 
lattice size 
it approaches the value $2$. We conjecture that $n_c=2$  is exact in the thermodynamic limit. }. The numerical results show that for
the values $0<n<2$ the equation (\ref{Renyi MI Potts})
is a very good approximation, as we can see for example in Fig.~10. The coefficient $\tilde{c}_n$ of the logarithmic term in 
(\ref{Renyi MI Potts}) is
obtained from the fitting of the model with 
$L=120$ sites is 
shown in  Fig.~11
and in the range $0.4<n<1.6$ surprisingly it follows the simple formula :

\begin{eqnarray}\label{c for HO} 
\tilde{c}_n=f(n)=1+4\frac{n-1}{10},\hspace{1cm}0.4<n<1.6 \quad.
\end{eqnarray}
This is the  red line in  Fig.~11. At $n=0$ we expect zero mutual information for our system, this means that based on
the symmetry $n\to2-n$ the coefficient for $n=2$ should be $\tilde{c}_2=2$. Finally one can conclude that for integer values of $n=0,1,2$ the coefficient
of the logarithm is
\begin{eqnarray}\label{c for HO integer} 
\tilde{c}_n=f(n)=n,\hspace{1cm}n=0,1,2.
\end{eqnarray}

\begin{figure} [htb] \label{fig11}
\centering
\includegraphics[width=0.35\textwidth]{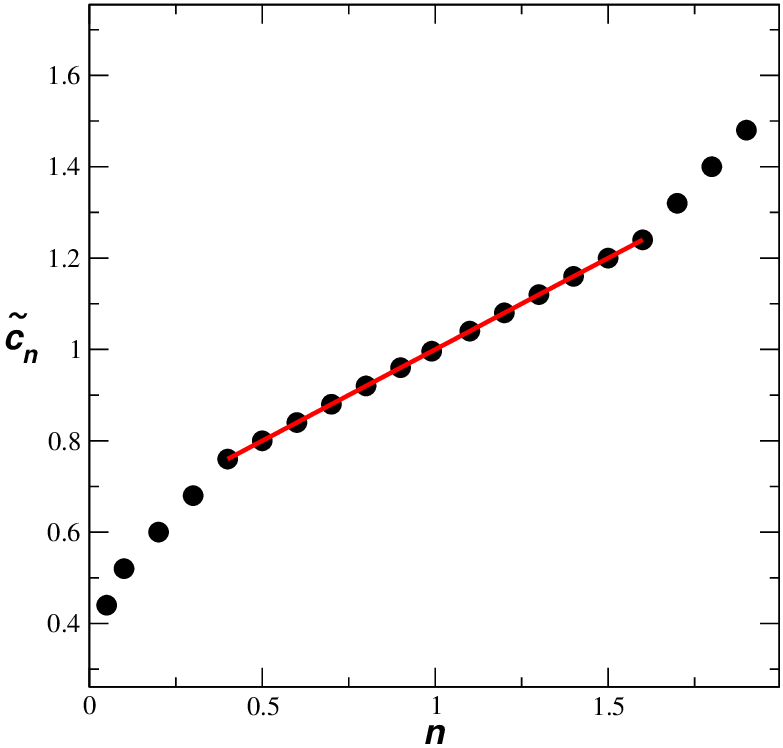}
\caption{ The coefficient of the logarithm  $\tilde{c}_n$ in the equation (\ref{Renyi MI Potts}).  
The lattice size $L=120$ and the coefficients $\tilde{c}_n$ were estimated  by using the subsystem 
sizes $\ell=3,5,...,{\mbox{Int}}[L/2]$. The red line is given by  Eq.~(\ref{c for HO}).} 
\end{figure}

\subsubsection{ The generalized mutual information of quantum spin chains with continuous symmetries} 

The Hamiltonian of the XXZ chain is defined as 
 \begin{eqnarray}\label{XXZ}
H_{\text{XXZ}}=-\sum_{i=1}^L(\sigma_j^x\sigma_{j+1}^x+\sigma_j^y\sigma_{j+1}^y+\Delta\sigma_j^z\sigma_{j+1}^z),
\end{eqnarray}
where $\sigma^x$, $\sigma^y$ and $\sigma^z$ are spin-$\frac{1}{2}$ Pauli
matrices and $\Delta$ is an anisotropy. 
The model is critical and conformal invariant for
 $-1\leq \Delta < 1$ with a constant central charge $c=1$, giving us  a good example
 to test the universality of our results with respect to the change of the anisotropy. 
 The long-distance critical fluctuations are 
ruled by a CFT with central charge $c=1$ described by a  compactified boson whose action is given by  
 \begin{eqnarray}\label{compactified Boson}
S=\frac{1}{8\pi}\int d^2x (\bigtriangledown\phi)^2,\hspace{1cm}\phi\equiv\phi+2\pi R,
\end{eqnarray}
where  the compactification radius depends upon the values of $\Delta$, 
namely:
\begin{equation} \label{R}
 R=\sqrt{\frac{2}{\pi}\arccos \Delta}. 
\end{equation}
 As it is shown in Fig.~12, in the
 $\sigma^z$ basis, the generalized mutual information 
$\tilde{I}_n(\ell,L-\ell)$   shows the logarithmic behavior given in (\ref{Renyi MI Potts}) only for $n<2$. 
This can be simply understood based on what we observed for the chain of harmonic oscillators. One can look to the Klein-Gordon field theory as a non-compactified
version of the action (\ref{compactified Boson}). Since we showed that in that case the generalized mutual information is not defined
beyond $n=2$ we expect the same behavior also in the compactified version. Note that in our numerical calculations one can actually derive 
spurious big numbers for the generalized mutual information even for $n>2$, but we expect all of them go to infinity in the thermodynamic limit.
 This behavior seems 
to be  independent of the anisotropy parameter $\Delta$.

The coefficient of the logarithm in 
  (\ref{Renyi MI Potts}) for $n<2$  is again given by \rf{coefficient}, as we can see 
in 
Fig.~13,  with a function $f(n)$ which fits to the results of the harmonic chain perfectly. We also considered the results in the case 
where the ground state wavefunction is expressed 
 in the $\sigma^x$ basis and,  except around $n=1$, the equation (\ref{Renyi MI Potts}) is not a good approximation.
\begin{figure} [htb] \label{fig12}
\centering
\includegraphics[width=0.35\textwidth]{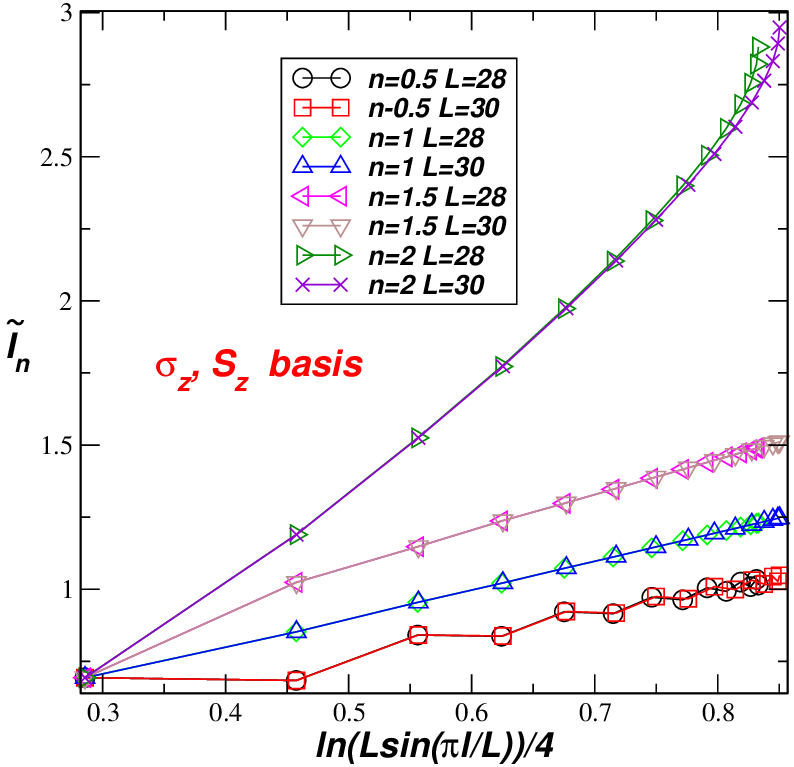}
\caption{ The generalized mutual information $\tilde{I}_n(\ell, L-\ell)$ of the periodic XXZ 
quantum chain with anisotropy $\Delta=-1/2$, as a function of 
 $\ln(\sin(\frac{\pi \ell}{L}))/4$. 
The ground-state wavefunction is in   the  basis where 
the $\sigma^z_i$ matrices are  diagonal ($\sigma^z$ basis). 
The results  are for lattice sizes $L=28$ and $L=30$ and give an idea of 
the finite-size corrections.}
\end{figure}
The 
second $U(1)$-symmetric model we considered is the spin-1 
Fateev-Zamolodchikov quantum  chain  whose Hamiltonian is given by \cite{fateev1}
\begin{eqnarray} \label{FZ}
H_{FZ} &=& \epsilon \sum_{i=1}^{L} \{ \sigma_i - (\sigma_i^z)^2
-2(\cos \gamma -1)(\sigma_i^{\perp}\sigma_{i}^z +
\sigma_i^z\sigma_{i}^{\perp}) \nonumber \\ &&-2\sin^2\gamma (\sigma_i^z
-(\sigma_i^z)^2 + 2(S_i^2)^2)\},
\end{eqnarray}
where $\vec{S} = (S^x,S^y,S^z)$ are spin-1 $SU(2)$  matrices,
$\sigma_i^z=S_i^zS_{i+1}^z$ and
 $\vec{S}_i\vec{S}_{i+1} = \sigma_i^{\perp} +\sigma_i^z$.
The model is antiferromagnetic  for $\epsilon =+1$ and ferromagnetic
for $\epsilon =-1$. It has
  a line of critical points
 ($0\leq\gamma \leq \frac{\pi}{2}$) with a quite distinct behavior in the
 antiferromagnetic ($\epsilon =+1$) and ferromagnetic
($\epsilon=-1$) cases.
The antiferromagnetic version of the model  is
governed by a CFT with central charge $c=\frac{3}{2}$ \cite{fateev2} while the ferromagnetic
one   is ruled by a $c=1$ CFT \cite{fateev3}. We calculated 
$\tilde{I}_n(\ell,L-\ell)$ in both critical regimes where $c=1$ and 
$c=\frac{3}{2}$, and for different values 
of the anisotropy. We  
  found a very similar pattern as that of   the XXZ quantum chain, 
as can be seen in  Fig.~13. The equation (\ref{Renyi MI Potts}) 
is valid for values of   $n<2$ and the coefficient of the logarithm follows (\ref{coefficient}) with a function  $f(n)$ which is quite similar 
 to the one   we found for the  
quantum harmonic oscillators and the XXZ chain. This shows an interesting universal pattern for critical chains with continuous $U(1)$ symmetry.

\begin{figure} [htb] \label{fig13}
\centering
\includegraphics[width=0.35\textwidth]{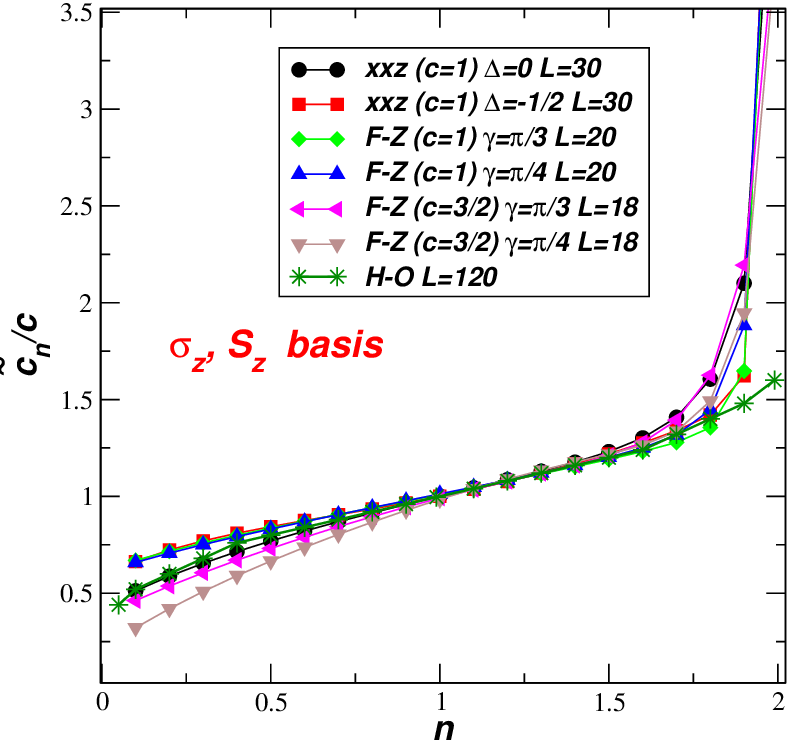}
\caption{The ratio $\tilde{c}_n/c$ of the coefficient of the logarithm in  equation (\ref{Renyi MI Potts}) with the central charge $c$ for the XXZ
 and for the spin-1 Fateev-Zamolodchikov quantum chains (F-Z). The XXZ 
(Fateev-Zamolodchikov) ground-state wavefunction are in the $\sigma^z$ 
($S^z$) basis.
The results for the XXZ are for the anisotropies $\Delta=0,-1/2$ and in 
the case of the Fateev-Zamolodchikov model their are for the 
couplings $\gamma=\pi/3,\pi/4$.  
  The lattice sizes of the models are shown and the coefficients $\tilde{c}_n$ were estimated  by using the subsystem 
sizes $\ell=4,5,...,L/2$.} 
\end{figure}

\section{Conclusions} 

In this paper we calculated the generalized mutual information 
$\tilde{I}_n(\ell,L-\ell)$, as defined in \rf{Renyi new}, for 
quantum chains describing the dynamics of quantum systems 
 with  
continuous or  discrete degrees of freedom.  
Most of our analysis was 
purely numerical due to the absence, at the moment, of suitable analytical 
methods to treat this problem. We considered several  integrable  quantum spin  
chains. These quantum chains either have a $Z(Q)$ symmetry (like 
the $Q$-state Potts model with $Q=2,3$ and 4, the Ashkin-Teller model, and 
the $Z(Q)$-parafermionic model with $Q=5,6,7$ and 8)  or 
a $U(1)$ symmetry (XXZ quantum chain and the spin-1 
Fateev-Zamolodchikov model). We also considered the discrete version of the 
Klein-Gordon field theory
 given by a set of  coupled harmonic oscillators. In this case we have a 
 continuum Hilbert space.
We observed  that 
by expressing the ground-state wavefunctions 
in general basis the obtained results are distinct. 
However, similarly as happens for the quantity $I_n$ given in 
\rf{Renyi2} (see \cite{AR2014}), our results on some special basis reveal  
some general features. 
These basis are the ones where the $S$ or $R$ operators are diagonal, for 
the models with $Z(Q)$ symmetry or the ones where $\sigma^z$ or $S^z$ are 
diagonal for the models 
with $U(1)$ symmetry. In a continuum field theory description of 
these quantum 
chains these basis are expected to be associated  to the boundaries that 
do not destroy the conformal invariance of the bulk underlying Euclidean 
conformal field theory, and for this reason we call them conformal 
basis \cite{AR2014}. Our results indicate that in these special basis the 
mutual information $\tilde{I}_n$ has the same kind of leading behavior 
with the subsystem size $\ell$ as we have in the R\'enyi entanglement entropy, 
namely $\tilde{I}_n(\ell,L-\ell) \sim \frac{\tilde{c}_n}{4} \ln(\frac{L}{\pi}\sin(\frac{\pi \ell}{L}))$, with a function $\tilde{c}_n=cf(n)$, with $f(1)=1$. 
Differently from the R\'enyi entanglement entropy where the equivalent 
function $f(n)$ is universal (for any model and any basis) in the case of 
$\tilde{I}_n$ our results indicate that the function $f(n)$ depends on 
the special basis chosen  to express the ground-state eigenfunction 
of  the particular model.
For the set of $Z(Q)$-symmetric  models 
we considered the function $f(n)$, for $n<4$, although different for the 
$S$ and $R$ basis are similar as the ones of  
the $Q$-state Potts chain ($Q=2,3,4$) and  
the parafermionic $Z(Q)$ quantum chains ($Q=5,6,7,8$). In the case of the 
Ashkin-Teller model our results indicate that $f(n)$, for $n>2$, also 
depends on the anisotropy $\Delta$ of the  model. On the other hand 
the models with 
continuum symmetry  showed a similar behavior only for $n<2$. 
 For $n>2$ we have strong evidences that most probably the generalized mutual information is not defined. It is quite interesting
 that in these cases one can understand most of the results by just studying simple short-range coupled harmonic oscillators.
 
In order to conclude we should mention that an analytical approach for 
the Shannon entropy or the Shannon mutual information ($I_1$ or $\tilde{I}_1$ 
in \rf{Renyi2} and \rf{Renyi new}) is a theoretical challenge. The analytical 
methods to treat this kind of problem normally use some sort of analytical 
continuation, in the parameter $n$, like the usual replica trick. The 
results we present showing the continuity of $\tilde{I}_n$ around $n=1$, 
differently from  what  happens with $I_n$, indicate that $\tilde{I}_n$ is probably 
more appropriate for an analytical treatment.

\textit{Acknowledgments}
This work was supported in part by FAPESP and CNPq (Brazilian agencies). We thank J. A. Hoyos, R. Pereira and V. Pasquier for useful discussions.

\section{Appendix: the relative entropy and the R\'enyi divergence} 
In this appendix we review the definitions of the  relative entropy and its generalization: the R\'enyi divergence. The relative entropy
is defined as the expectation of the difference between the logarithm of the 
two distribution of probabilities $p$ and $q$, from the point of view of 
the distribution $p$, i. e.,
\begin{eqnarray}\label{relative entropy}
D(p\parallel q)=\sum_i p_i\ln\frac{p_i}{q_i}.
\end{eqnarray}
It can be considered as a measure of the difference between the two distributions $p$ and $q$. Although it is not a symmetric quantity  it helps us to 
define the mutual information 
of the subsets $X$ ans $Y$ of the system as follows
\begin{eqnarray}\label{mutual information from relative entropy}
I(X,Y)=D(p(X,Y)\parallel p(X)p(Y)).
\end{eqnarray}
In words,  the mutual information between two parts of a system is just the relative entropy between the distribution probability 
for the whole system and the product of the probability distributions of the different parts. It tells how much the different parts are correlated.
The natural generalization of the relative entropy is the R\'enyi divergence and can be defined (see \cite{Principe} for example), as 
\begin{eqnarray}\label{Renyi divergence}
D_n(p\parallel q)=\frac{1}{n-1}\ln\sum_i p_i^nq_i^{1-n}.
\end{eqnarray}
It has the following properties: for $n>0$ we have $D_n(p\parallel q)\neq 0$ and if $p=q$ then we have $D_n(p\parallel q)=0$. The especial case $n\to 1$
gives the usual relative entropy. We also define the $n=0$ case by:
\begin{eqnarray}\label{mutual information from Renyi divergence}
D_0(p\parallel q)=-\ln q({i|p_i>0}).
\end{eqnarray}
It is worth mentioning that using the above definition $D_0(p\parallel q)$ is not zero except when for all $i$'s for which $q_i>0$ also $p_i>0$ holds.

Another important property is the following (see \cite{vanErven} and references therein):
\newline
\newline
\emph{Theorem}: the R\'enyi divergence is a continuous  and nondecreasing 
function of the parameter  $n$.

Comparing  
\rf{Renyi divergence} with 
\rf{mutual information from relative entropy} and 
\rf{relative entropy} the natural definition of the generalized mutual 
information is 
\begin{eqnarray}\label{mutual information from Renyi divergence}
\tilde{I}_n(X,Y)=D_n(p(X,Y)\parallel p(X)p(Y)).
\end{eqnarray}
The above definition is different from $I_n(\ell,L)$, as 
given by \rf{Renyi2}, and has been  frequently used in different areas of information science.
 
\end{document}